\documentclass[aps,prb,twocolumn,showpacs,amsmath,floatfix]{revtex4}
\usepackage{amsmath,graphics,epsfig,color,verbatim,ulem}
\usepackage{dcolumn,multirow}
\usepackage{subfigure}
\pdfoutput=1

\begin{document}

\title{Strong coupling of Jahn-Teller distortion to oxygen-octahedron rotation and 
functional properties in epitaxially-strained orthorhombic LaMnO$_3$}

\author{Jun Hee Lee$^{1}$}
\email{jhlee@physics.rutgers.edu}
\author{Kris T. Delaney$^{2}$}
\author{Eric Bousquet$^{3}$}
\author{Nicola A. Spaldin$^{3}$}
\author{Karin M. Rabe$^{1}$}

\affiliation{
$^1$Department of Physics and Astronomy, Rutgers University, Piscataway, New Jersey 08854-8019, USA\\
$^2$Materials Research Laboratory, University of California, Santa Barbara, California 93106-5121, USA\\
$^3$Department of Materials, ETH Zurich, Wolfgang-Pauli Strasse 27, CH-8093 Zurich, Switzerland
}
\marginparwidth 2.7in

\marginparsep 0.5in

\begin{abstract}

First-principles calculations reveal a large cooperative coupling of Jahn-Teller (JT) distortion to oxygen-octahedron rotations 
in perovskite LaMnO$_3$. 
The combination of the two distortions is responsible for 
stabilizing the strongly orthorhombic $A$-AFM insulating ($I$) $Pbnm$ ground state relative 
to a metallic ferromagnetic (FM-$M$) phase. 
However, epitaxial strain due to coherent matching to a crystalline substrate can change the relative stability of the two states.
In particular, coherent matching to a square-lattice substrate favors 
the less orthorhombic FM-$M$ phase, with the $A$-AFM phase stabilized at higher values of tensile epitaxial strain 
due to its larger volume per formula unit, resulting in a coupled magnetic and metal-insulator transition 
at a critical strain close to 1\%. 
At the phase boundary, colossal magneto-resistance is expected. 
Tensile epitaxial strain enhances the JT distortion and opens the band gap in the $A$-AFM-$I$ $c$-$Pbnm$ phase, 
offering the opportunity for band-gap engineering. 
Compressive epitaxial strain induces an orientational transition within the FM-$M$ phase from $c$-$Pbnm$ to $ab$-$Pbnm$ 
with a change in the direction of the magnetic easy axis relative to the substrate, yielding strain-controlled magnetization at
the phase boundary. 
The strong couplings between the JT distortion, the oxygen-octahedron rotations 
and the magnetic and electronic properties, and associated functional behavior,
motivate interest in other orthorhombic $Pbnm$ perovskites with large JT distortions, which should also 
exhibit a rich variety of coupled magnetic, 
structural and electronic phase transitions driven by epitaxial strain. 

\end{abstract}


\maketitle

(La,$M$)MnO$_3$ ($M$=Ca,Pr,Sr,Ba) has been of great interest due to the couplings among structure, magnetic and orbital orderings and the
resulting functional properties, such as colossal magneto-resistance (CMR) \cite{CMR1,CMR2,CMR3}. 
As a result, there have been many experimental \cite{exp1,exp2,easy-axis1,exp4} 
and theoretical \cite{theo1,easy-axis2,failure,Leonov,Terakura10} studies of the end-member compound LaMnO$_3$ (LaMnO$_3$).
The observed sequence of phases with decreasing temperature is as follows\cite{exp4,JT-Tc}.
At very high temperature (above 1010 K), LaMnO$_3$ has a rhombohedral $R\overline{3}c$ structure, produced 
by rotation of the oxygen octahedron network of the ideal perovskite structure. 
From 1010 K down to 750 K, it has an orthorhombic $Pbnm$ structure produced by a different pattern of octahedral rotations. 
At 750 K, an orbital-order transition occurs by a cooperative Jahn-Teller transition \cite{Billinge}; however, this distortion 
does not break the symmetry of the $Pbnm$ structure. Finally, a magnetic transition 
to an $A$-type antiferromagnetic ($A$-AFM) ordering (ferromagnetic alignment in the $xy$ planes, with spin direction alternating 
from plane to plane) occurs at 135 K. 

It proves to be challenging to reproduce the observed $A$-AFM ground state of LaMnO$_3$, 
with its half-filled $e_g$ level, from first principles. 
If the structural parameters are fixed 
to the experimental values, the correct magnetic ordering is obtained with density-functional-theory (DFT) methods. 
However, full optimization of the structure within DFT gives a competing FM-$M$ phase as the ground state, 
and to obtain the correct ground state requires use of a generalized gradient form of the density functional 
and inclusion of a Hubbard $U$ \cite{Terakura10}.

More generally, various couplings in complex oxides produce a rich variety of 
distinct low-energy alternative phases with distinct structure and magnetic ordering \cite{Rondinelli}, 
as demonstrated in EuTiO$_3$ \cite{Rabe06,EuTiO2}, 
SrMnO$_3$ \cite{ours-SMO}, and SrCoO$_3$ \cite{ours-SCO}. 
Certain phases are favored by epitaxial strain, leading to epitaxial-strain-induced 
coupled first-order phase transitions with associated functional properties at the phase boundaries.
Similarly, previous first-principles studies of LaMnO$_3$ demonstrate electronic and magnetic phase transitions 
driven by uniaxial strain \cite{uniax1,uniax2,uniax3,uniax4}. 
However, oxygen-octahedron rotation distortions and their coupling to JT distortions \cite{coupling}
were not considered in the latter studies, and it is expected that
including these distortions could give rise to additional phases and provide additional coupling to epitaxial strain.

In this paper, we report the phase sequence for LaMnO$_3$ under epitaxial strain, obtained from first-principles calculations, 
with the combination of oxygen-octahedron rotations and Jahn-Teller distortions. 
Compressive epitaxial strain on a square-lattice substrate 
is found to favor the less-orthorhombic, higher-density FM-$M$ phase relative to the AFM-I phase, 
with a coupled magnetic insulator-metal transition at a critical strain close to 1\%.
At the phase boundary, we expect functional properties, including large magnetoresistance.
In the epitaxially-strained AFM phase, increasing tensile strain enhances the JT distortion and increases the band gap; this tunability 
could be used in tailoring the catalytic effectiveness of LaMnO$_3$ for energy applications including fuel cells \cite{ORR11}.
In the epitaxially-strained FM phase, higher compressive strain induces an orientational transition resulting
in change of the direction of the ferromagnetic easy axis, yielding strain-controlled magnetization at
the phase boundary. 

\renewcommand{\arraystretch}{1.2}
\begin{table}[!]
\caption{Comparison between the results of our GGA+$U$ calculation ($U_{\rm eff}$=1.7 eV), 
a previous GGA+$U$ calculation ($U_{\rm eff}$=2.0 eV)\cite{Terakura10}, and 
experiment for the ground-state $Pbnm$ $A$-AFM phase of LaMnO$_3$. 
The Jahn-Teller distortion magnitudes Q$_2$ and Q$_3$ (defined in Ref. \onlinecite{Terakura10}) are in a.u., the oxygen-octahedron 
rotation angles $\theta_{in}$ (for the rotation around [110]) and $\theta_{out}$ (for the rotation around [001]) in degree($^\circ$), 
Mn local magnetic moment $m$ in $\mu_{\rm B}$/f.u., intersite magnetic exchange couplings $J_{\rm in}$ (in the $ab$ plane) and $J_{\rm out}$ (along the $c$ direction) in meV/f.u., and direct ($\Delta E_d$) and indirect ($\Delta E_i$) band gaps in eV. 
Experimental data were taken from Ref. \onlinecite{mag} (structure, magnetic moment), 
Ref. \onlinecite{easy-axis1} (intersite exchange coupling), 
Ref. \onlinecite{gap2} (direct gap), and Ref. \onlinecite{gap1} (indirect gap).}

\begin{ruledtabular}
\begin{tabular}{cccccccccc}
      &          \multicolumn{4}{c}{structure}              &   \multicolumn{3}{c}{magnetism}  & \multicolumn{2}{c} band gaps \\ 
      & Q$_2$ & Q$_3$ & $\theta_{\rm in}$ & $\theta_{\rm out}$ & m  & $J_{\rm in}$ & $J_{\rm out}$ &    $\Delta E_d$      & $\Delta E_i$     \\ 
\hline
Cal.      & 0.14& 0.83&    152.2  &     152.4  & 3.63 & 0.80  &  -0.47 &     1.1   &   0.83 \\
Cal.\cite{Terakura10} & 0.12& 0.68&    153.1  &    152.8  & 3.49 &   &  -0.65 &     1.2   &   0.90 \\
Exp.&0.14&0.78&154.3&156.7&3.7$\pm$0.1& 0.83 & -0.58 & 1.1 & 0.24 \\ 
\end{tabular}
\end{ruledtabular}
\label{reproduction}
\end{table}

\begin{table}
\caption{Lattice parameters (\AA) and Wyckoff positions of bulk LaMnO$_3$ from experiment\cite{mag} and GGA+$U$ calculations (this work).}
\begin{ruledtabular}
\begin{tabular}{ccccc}
          &   & $A$-AFM (exp.)& $A$-AFM (cal.) & FM (cal.)\\
\hline
     a     &    & 5.532 & 5.571 & 5.566  \\
     b     &    & 5.742 & 5.857 & 5.616  \\
     c     &    & 7.668 & 7.703 & 7.901  \\
 La   (4c)& x & 0.510 & 0.511 & 0.508  \\
          & y & 0.451 & 0.441 & 0.463  \\
 O$_2$(8d)& x & 0.724 & 0.721 & 0.717  \\
          & y & 0.691 & 0.687 & 0.713  \\
          & z & 0.039 & 0.043 & 0.041  \\
 O$_1$(4c)& x & 0.430 & 0.417 & 0.423  \\
          & y & 0.014 & 0.017 & 0.017  \\
\end{tabular}
\end{ruledtabular}
\label{Wyckoff}
\end{table}
\renewcommand{\arraystretch}{1.0}

First-principles calculations were performed using density-functional 
theory within the
generalized gradient approximation plus Hubbard $U$ (GGA+$U$) method \cite{GGAU}
with the Perdew-Becke-Erzenhof parameterization \cite{PBE}
as implemented in
the $Vienna$ $Ab$ $Initio$ $Simulation$ $Package$ 
(VASP-5.2)~\cite{Kresse2,Kresse3}. 
We use the Liechtenstein~\cite{Liech} implementation 
with on-site Coulomb interaction $U$=2.7 eV
and on-site exchange interaction $J_H$=1.0 eV 
to treat the localized 3$d$ electron states in Mn; this choice of $U$ is close to that chosen in previous work \cite{Terakura10}.
The projector augmented wave (PAW) potentials \cite{Kresse1} explicitly
include 11 valence electrons for La ($5s^2 5p^6 6s^2 5d^1$), 13 for Mn (3$p^6$3$d^5$4$s^2$), 
and 6 for oxygen (2$s^2$2$p^4$). 
We used a $\sqrt 2 \times \sqrt 2 \times 2$ perovskite supercell, 
a $4 \times 4 \times 4$ Monkhorst-Pack $k$-point mesh, and a 500 eV plane-wave cutoff 
for total energy calculations and structural optimization using Hellmann-Feynman forces. 
Spin-orbit coupling was included in calculations of the magnetocrystalline anisotropy. 

In this work, epitaxial strain is defined
as the in-plane strain produced by coherent matching of the material 
to a square-lattice substrate with lattice parameter $a$, quantified as $(a-a_0)/a_0$ with $a_0$=3.976\AA,
the cube root of the computed volume per formula unit of the relaxed $A$-AFM 
$Pbnm$ structure. 
To study the effects of epitaxial strain, we performed
``strained-bulk'' calculations \cite{strain}, in which the structural parameters ($c$ lattice, ionic positions) 
of the bulk periodic supercells are optimized subject
to the constraint that the two in-plane lattice vectors which define the matching plane are fixed to produce the specifed square lattice. 
At each value of the strain, we considered FM, $A$-AFM, $C$-AFM and $G$-AFM magnetic ordering for the epitaxially-constrained 
$R\overline{3}c$ structure and for the two distinct orientations of the epitaxially-constrained $Pbnm$ structure, 
which we refer to as $c-Pbnm$ and $ab-Pbnm$, as in previous work on SrRuO$_3$ \cite{sro} and CaTiO$_3$ \cite{ab-Pbnm}.
As we found that $C$- or $G$-type and $R\overline{3}c$ phases do not appear as the lowest energy phases 
in the strain range considered, they are not discussed further in this paper. 
In addition, we also checked for polar instability of the $Pbnm$ phases along [001] and [110] at -4 \% and +4 \%, 
respectively, but the nonpolar phases were found to be stable, and thus polar distortions were not considered further.

\begin{figure}
\begin{center}
\includegraphics[width=8.8cm,trim=0mm 4mm 0mm 0mm]{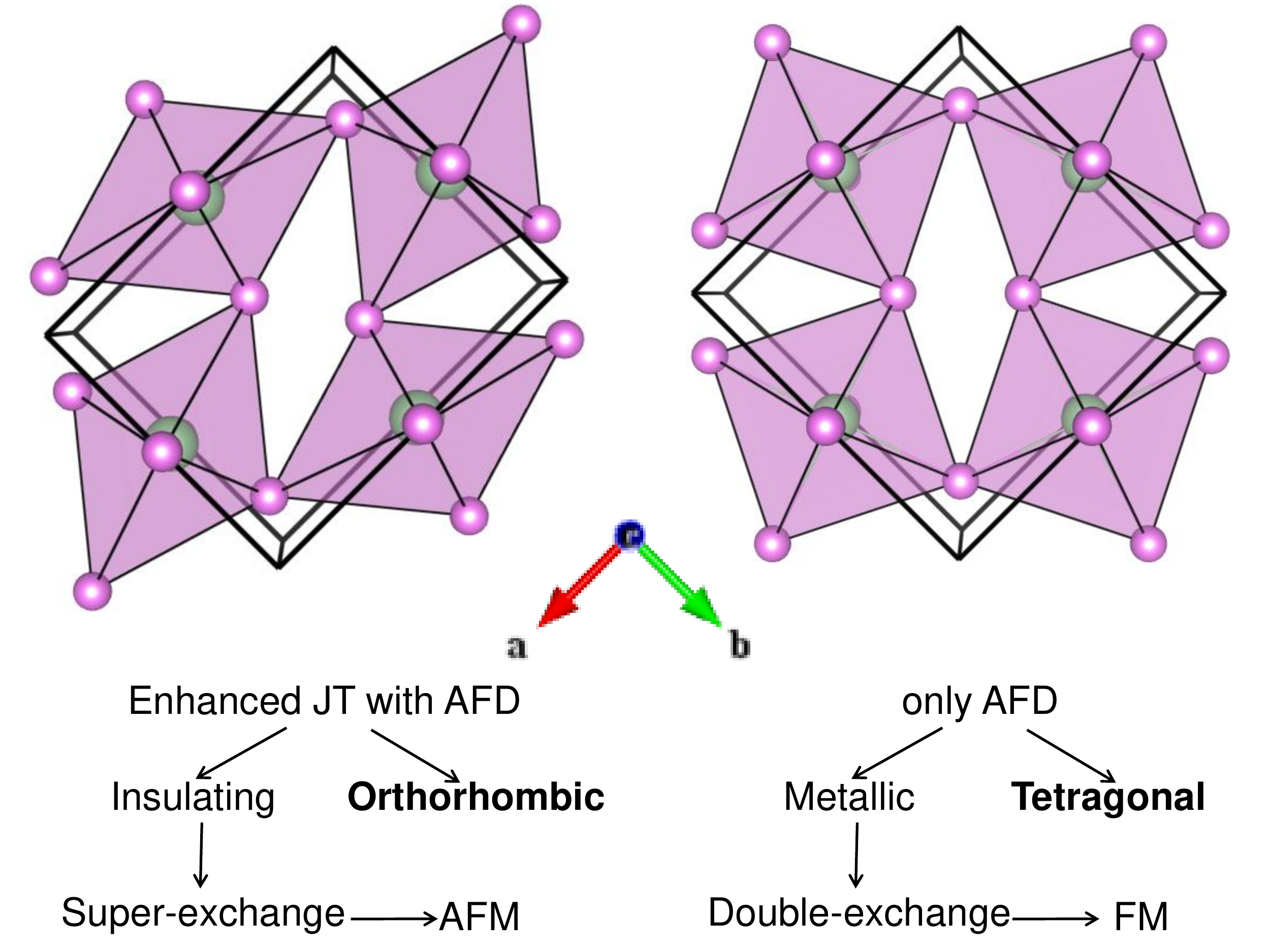}
\end{center}
\caption{Comparison of bulk AFM orthorhombic (left) and bulk FM tetragonal (right) structures.}
\label{comparison}
\end{figure}

\renewcommand{\arraystretch}{1.3}
\begin{table}
\caption{Properties of the ground-state and low-energy competing bulk phases in LaMnO$_3$ calculated in this work.} 
\begin{ruledtabular}
\begin{tabular}{ccc}
                                                   &  Bulk phase            & Alternative phase      \\ 
\hline 
 Magnetic order                                    & $A$-AFM                & FM                     \\
 Relative energy                                   &  0 meV/f.u.            & 13 meV/f.u.       \\
 Electronic property                               &  Insulating            & Half-metallic          \\
 JT magnitude (Q$_3$)                              &  0.83 a.u.             &  0.07 a.u.             \\
Rotation angles ($\theta_{\rm R}, \theta_{\rm M}$) & 14$^\circ$, 10$^\circ$ & 13$^\circ$, 8$^\circ$  \\
 Volume/fu                                         &  62.84 \AA$^3$         &   61.74 \AA$^3$        \\
Orthorhombicity($\frac{b}{a}$-1, $\frac{c}{\sqrt2a}$-1) & 5.2\%, -2.3\%     &    0.90\%, 0.29\%      \\
\end{tabular}
\end{ruledtabular}
\label{alternative}
\end{table}

\renewcommand{\arraystretch}{1.0}

In Tables ~\ref{reproduction} and ~\ref{Wyckoff}, we report the results of the first-principles calculations 
for the orthorhombic $Pbnm$ structure with $A$-AFM ordering. 
Consistent with experiment and previous first-principles results, we find this to be the ground state, 
with good agreement for the structural parameters and other properties, including the magnitude of the Jahn-Teller distortions, 
the oxygen octahedron rotation angles, the local magnetic moment of Mn, the exchange coupling, and the direct band gap. 
The indirect gap (0.83 eV) involves a conduction-band minimum at $\Gamma$ and valence-band maximum
near $R$.
This value is close to the previous theoretical result (0.90 eV) \cite{Terakura10}, though 
substantially higher than the experimental value (0.24 eV) extracted from transport measurements \cite{gap1}. 

\begin{figure}\begin{center}\includegraphics[width=10.5cm,trim=36mm 10mm 2mm 1mm]{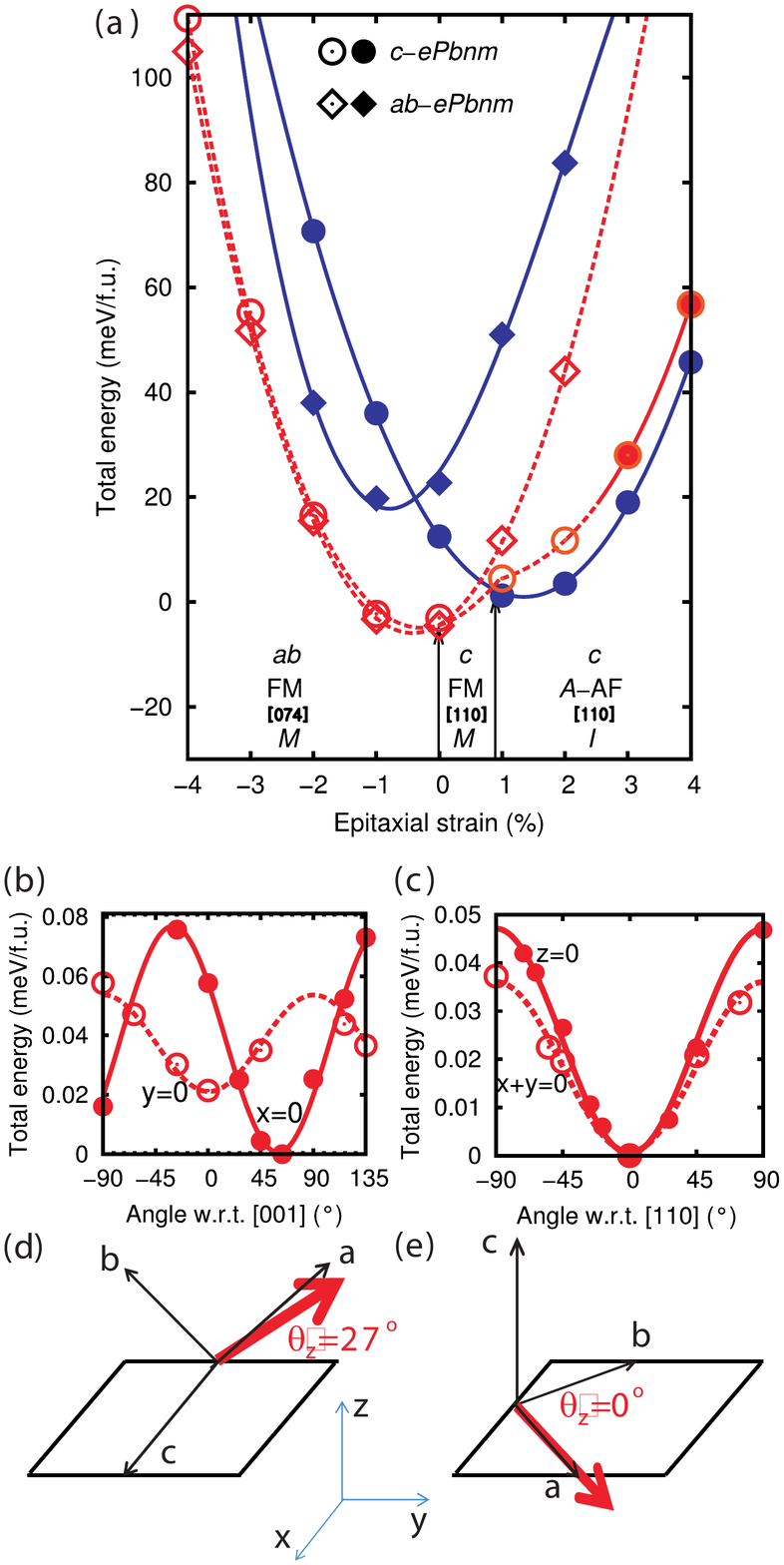}
\end{center}
\caption{(Color online) 
(a) Computed total energies for relaxed $c$-$Pbnm$ and $ab$-$Pbnm$ structures 
with different magnetic orderings at each strain in LaMnO$_3$. 
Thick lines represent insulating character and dotted lines, metallic. Blue (filled) symbols represent $A$-AFM and 
red (empty) symbols represent FM ordering. The lowest energy phase at each strain is specified along the bottom of the
plot, separated by vertical lines representing the phase boundaries; the subscripts give the direction of the magnetic easy axis. 
In (b) and (c), we show the magnetocrystalline anisotropy energies computed in various planes of FM $ab$-$Pbnm$ 
and FM $c$-$Pbnm$, respectively, as labeled on the figure; 
the zero of energy is chosen as the minimum energy obtained in each panel.
In (d) and (e), or FM $ab$-$Pbnm$ and FM $c$-$Pbnm$ respectively, we show the lattice vectors relative 
to the substrate plane with the direction of the magnetic easy axis represented as a thick red arrow and labeled 
with the value of the angle it makes with the substrate plane.}
\label{energy}
\end{figure}

In Table~\ref{Wyckoff}, we also report the first-principles structural parameters for the $Pbnm$ FM-$M$ phase, which is 
half-metallic with a gap of 3 eV in the semiconducting spin channel.
In Table~\ref{alternative} and Figure~\ref{comparison} we compare the structure and properties of this phase 
with the ground-state $Pbnm$ $A$-AFM-$I$ phase.
The most striking difference is that the $A$-AFM phase is strongly orthorhombic, while the FM-$M$ phase is nearly tetragonal. 
As we illustrate in Figure~\ref{comparison} and will discuss further below, 
the degree of orthorhombicity is related
to the magnitude of the JT distortions: large in the $A$-AFM phase and almost negligible in the FM-$M$ phase.
This relationship is corroborated by the low orthorhombicity of 
other $Pbnm$ perovskites that have rotational distortions similar to those of LaMnO$_3$ but negligible JT distortions,
such as LaScO$_3$, with ($\frac{b}{a}$-1, $\frac{c}{\sqrt2a}$-1) = (2.4\%, 0.76\%), and SrZrO$_3$, with (0.21\%, 0.02\%) \cite{Pbnm}. 
Another important point evident in Table~\ref{alternative} is that the energy difference between AFM-$I$ and FM-$M$ is small, 
and thus that the FM-$M$ phase could be stabilized by an appropriate perturbation.  
The large difference in orthorhombicity suggests that epitaxial growth on a square-lattice substrate could be effective.
Matching to a square lattice would force distortion of the orthorhombic AFM-$I$ phase, costing elastic energy, 
while the nearly tetragonal FM-$M$ phase can match to a square lattice with little elastic energy penalty for the shape change.
Furthermore, the FM-$M$ phase, with a smaller volume per formula unit, will be favored by compressive strain.

\begin{figure}[t!]\begin{center}\includegraphics[width=11cm,trim=31mm 12mm 2mm 2mm]{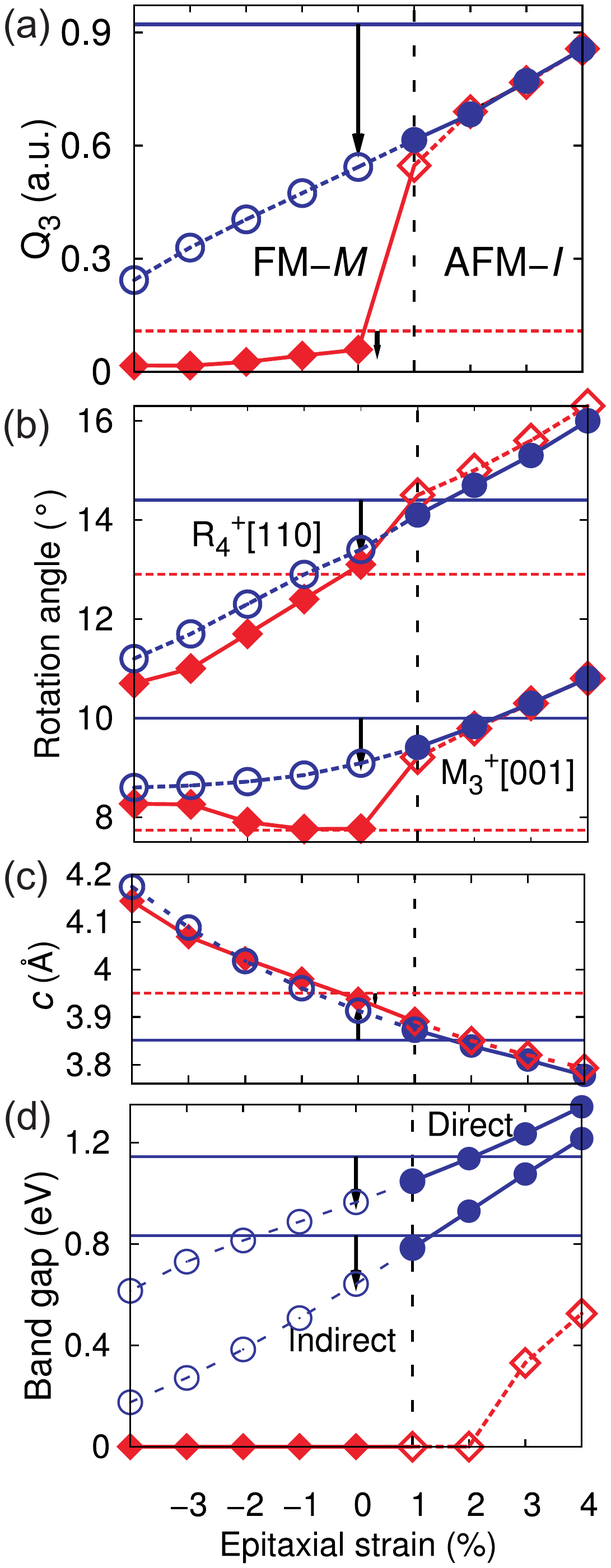}\end{center}\caption{(Color online) 
Epitaxial strain dependence for $A$-AFM ordering (circles) and FM ordering (diamonds) of
(a) the JT distortion amplitude; (b) rotation angles for R and M distortions; 
(c) the $c$ lattice parameter and (d) the direct and indirect band gaps. 
Horizontal blue (red) lines represent values in the bulk $A$-AFM ground state (FM alternative state) for comparison.
Arrows highlight the differences in properties between bulk and 0 \% strained phase. 
The symbol corresponding to the lowest-energy state at a given strain is filled and the symbols representing 
the higher-energy state at the same strain are open.}
\label{property}
\end{figure}

Figure~\ref{energy} shows total energy results for $A$-AFM and FM phases under a square-lattice epitaxial-strain constraint ranging from 
-4 \% to +4 \% in $c$-$Pbnm$ and $ab$-$Pbnm$ \cite{ab-Pbnm} structures.
As expected, the FM-$M$ phase is favored for compressive strain, while for tensile strain the AFM-$I$ phase is stable. 
The figure shows that the FM-$M$ phase is favored over the $A$-AFM phase at 0\% strain; this is in fact 
not inconsistent with the prediction of the $A$-AFM phase as the ground state, 
since on a square substrate, any value of epitaxial strain requires substantial distortion of this strongly orthorhombic phase.
With increasing compressive strain, there is an orientational transition in which the matching plane 
for the FM-$M$ $Pbnm$ phase changes from the (001) plane ($c-Pbnm$) to the (110) plane ($ab-Pbnm$), though it should be noted that the
energy difference between the two phases is quite small since the FM-$M$ phase is close to cubic.
With increasing tensile strain, FM $Pbnm$ exhibits an isosymmetric JT transition at 1 \% (Figure \ref{property}(a)) 
and a subsequent metal-insulator transition at around 3 \%, as shown in Figure ~\ref{energy} and ~\ref{property}(d).
For $A$-AFM, compressive strain strongly favors $ab$-$ePbnm$ because the short $c$ lattice vector
($|\vec{c}|/$$\sqrt2$=5.447\AA~in bulk, compared to $|\vec{b}|$=5.857\AA~and $|\vec{b}|$=5.571\AA) is in the substrate matching plane.
In the high-tensile strain phase, the structures for the FM and AFM orderings 
are very similar (Figure \ref{property}(a-c)), reflecting a weak spin-lattice coupling.


At the critical strain, which is just under 1\%, we expect a first-order coupled magnetic and metal-insulator transition. 
For a system in the vicinity of the phase boundary, large magnetic-electronic responses 
are predicted as the result of the 
possibility of switching two phases with quite different magnetic ordering and band gap \cite{Rabe06,ours-SMO,ours-SCO,CMO}.
In particular, application of a magnetic field to the AFM-$I$ phase just at the phase boundary could drive the system to the FM-$M$ phase, 
with a jump in electrical conductivity. This would produce a colossal magnetoresistance effect, of particular interest as it would be in a pure compound rather than in a mixed system with cation substitution \cite{CMR1,CMR2,CMR3}.

Close to 0\% strain, we find a phase transition due to the change in the lowest energy matching
plane for both the FM and AFM $Pbnm$ phases, which changes the orientation of the $Pbnm$ lattice vectors
relative to the substrate, and leads to a lowered symmetry in the $ab-Pbnm$ orientation. 
This change in orientation and symmetry lowering results in a change of the
magnetic easy axis direction relative to the substrate.
In bulk $A$-AFM $Pbnm$ LaMnO$_3$, the magnetic easy axis 
is along [110] \cite{easy-axis1,easy-axis2,easy-axis3} 
and the measured magnetocrystalline anisotropy energy is 0.66 meV/fu \cite{easy-axis3}, 
in good agreement with our computed value of 0.50 meV/fu. 
Our computations for the magnetocrystalline anisotropy energy of the FM-$M$ $ab-Pbnm$ and $c-Pbnm$ phases
are shown in Figure~\ref{energy}(b) and (c), respectively.
For FM $c$-$Pbnm$, the easy axis is also along $\vec{a}$, so that it lies in the plane of the substrate. 
In the $ab$-$Pbnm$ orientation, $\vec{a}$ is tilted out of the substrate plane, and further, as
a result of the lower symmetry, the easy axis is slightly tilted from $\vec{a}$ (by 18$^\circ$)
forming an angle of 27$^\circ$ with the substrate. 
A discontinuous change in the direction of the FM easy axis, from [110] to [074], is thus expected  at 0 \% strain. 
In the vicinity of the critical strain, it should be possible to switch from one orientation to the other
by application and removal of a uniaxial stress along the normal to the substrate, yielding a
strain-controlled magnet. 

\begin{figure}
\begin{center}
\includegraphics[width=8.6cm,trim=0mm 0mm 0mm 0mm]{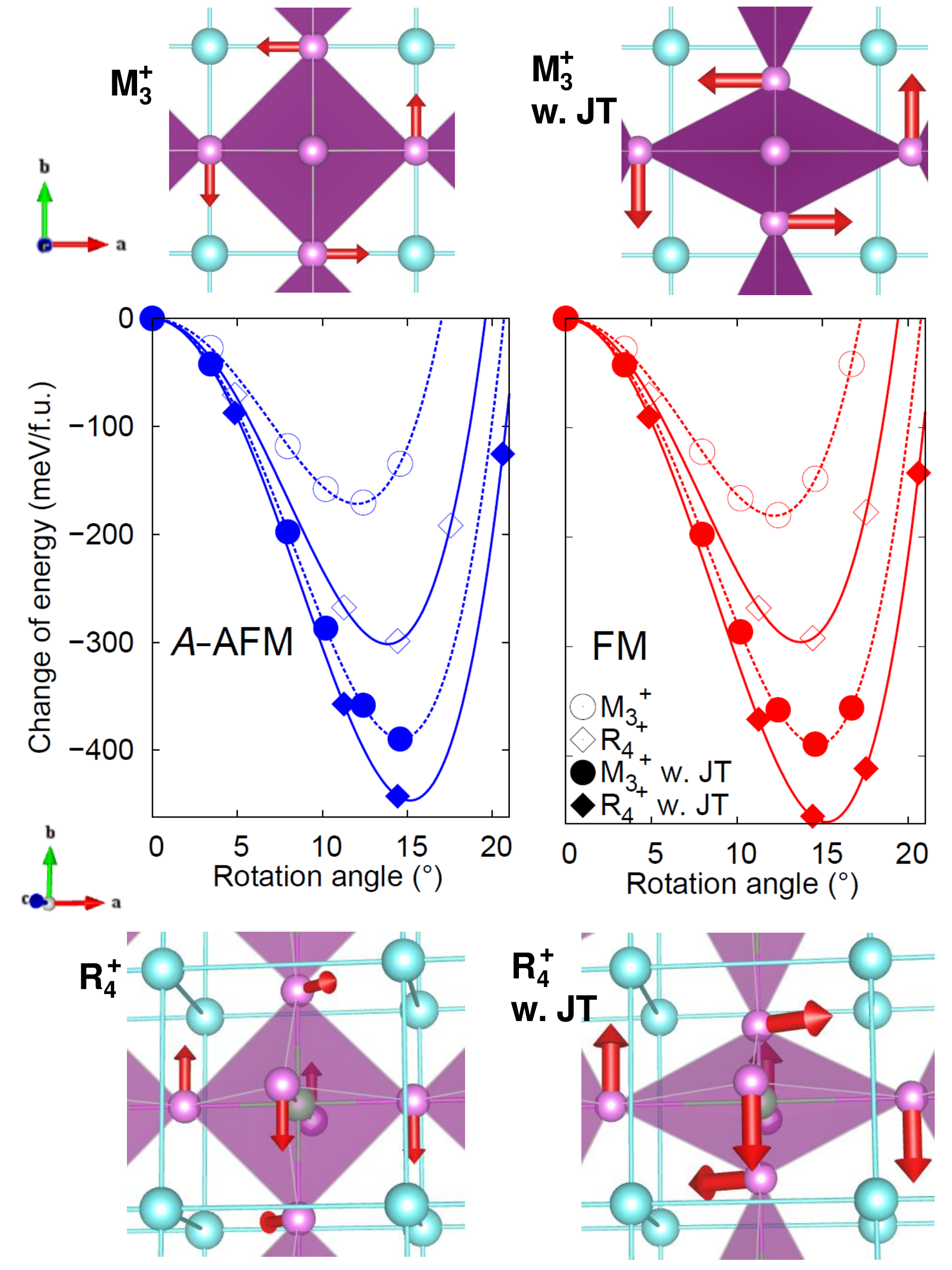}
\end{center}   
\caption{(Color online) The dependences of total energies on the amplitudes of 
oxygen-octahedron-rotation distortions M$_3^+$[001] (circles) 
and R$_4^+$[110] (diamond)
of cubic LaMnO$_3$ ($a_{\rm 0}$=3.976\AA) are shown for $A$-AFM ordering (left) and FM ordering (right).
Open symbols represent pure rotations and filled symbols represent rotations in the presence of 
a nonzero Jahn-Teller distortion (Q$_3$=0.926 a.u.). 
For each curve, the zero of energy is taken at zero rotation.
Thick lines represent insulating systems and dotted lines represent metallic systems. 
Total energies are set to zero when there are no oxygen-octahedron rotations.}
\label{rotation}
\end{figure}

To investigate the origin of the competing FM-M and A-AFM-I states and their dependence on epitaxial strain, 
we have performed first-principles calculations to consider the energetics of the JT distortion 
and the oxygen octahedron distortions, taken individually, and the coupling between them in cubic reference phases 
with FM and A-AFM ordering. 
First, we compute the magnetic ordering energy in the cubic reference structure with $a_{\rm 0}$=3.976 \AA, and find that
FM is favored over the $A$-AFM phase by 98 meV/fu. 
Next, we consider pure rotations of oxygen octahedra in the cubic reference structure. 
Figure~\ref{rotation} shows total energy changes with respect to the two oxygen-octahedron rotation modes, M$_3^+$[001] and R$_4^+$[110], that appear in the orthorhombic $Pbnm$ structure \cite{ISODISTORT}.
The large energy gain for the R$_4^+$[110] mode, in particular, is consistent 
with observation of a high-temperature $R\overline{3}c$ phase, which is generated by this distortion.
For both modes, the energy as a function of angle is almost identical for $A$-AFM ordering and FM ordering, 
so that the spin-phonon coupling for the rotation modes is found to be small. 
We infer that rotations are not responsible for driving the FM cubic system to the ground state $A$-AFM phase because 
the energy gains from the rotations in FM and in $A$-AFM are the same. 
Furthermore, we see that the states with pure rotational distortions are all metallic, and thus that
the oxygen-octahedron rotations are not responsible for the insulating character of bulk $Pbnm$ LaMnO$_3$. 

\begin{figure} \begin{center} \includegraphics[width=8.5cm,trim=0mm 0mm 0mm 0mm]{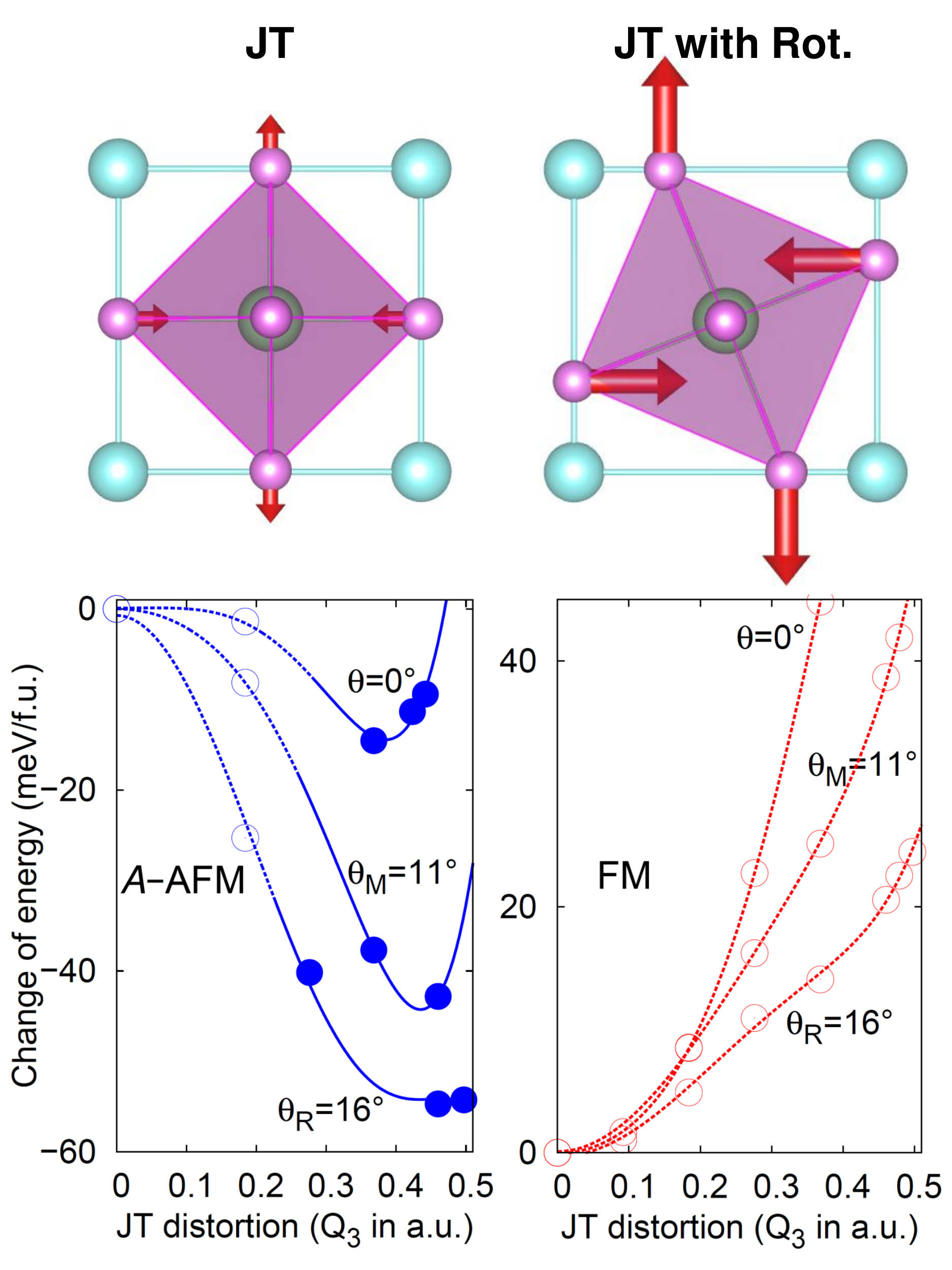} \end{center} \caption{(Color online) 
The geometrical interpretation of how rotations lower the energy cost of the Jahn Teller distortion
is illustrated in the top half of the figure.
In the bottom half, the dependences of total energies on the amplitude of the Jahn-Teller distortions (M$_2^+$[001]) 
for zero and two nonzero oxygen octahedron rotation distortions are shown for $A$-AFM ordering (left) and FM ordering (right).
For each curve, the zero of energy is taken at zero Jahn-Teller distortion amplitude. As in Figure~\ref{rotation}, 
thick lines with filled circles represent insulating systems and dotted lines with blank circles, metallic.}
\label{JT}
\end{figure}
Next, we consider the JT (M$_2^+$[001]) distortion.
Figure~\ref{JT} shows the total energy change with respect to this distortion for the $A$-AFM and FM phases.
For FM ordering, the pure distortion is stable.
For $A$-AFM ordering, the pure distortion is unstable, demonstrating a huge spin-phonon coupling. However, the energy gain of 13 meV/fu is
not enough to stabilize the $A$-AFM ordering relative to the FM ordering, 90 meV/fu lower, and is also small relative to that for the pure
R$_4^+$ rotation (400 meV/fu), consistent with the fact that the JT transition occurs at a temperature well below that at which the
rotational distortion becomes favorable. The JT distortion is quite effective at opening the band gap in the presence of $A$-AFM ordering,
with a metal-insulator transition at $Q_3 \approx$ 0.3 a.u., so that this distorted structure is insulating.

We now consider the coupling between the rotations and the JT distortions. From the first principles results in Figures ~\ref{rotation} and ~\ref{JT}, we can see that the couplings between the rotations and the JT distortion are quite
strong.
Specifically, the energy gains from the M and R rotations are noticeably increased in the presence of a nonzero JT M$_2^+$[001]
distortion, and the energy gain for the JT distortion is increased by several times in the presence of the oxygen-octahedron rotations.
This cooperative coupling can be understood from consideration of the geometry of the crystal structure as illustrated in Figure~\ref{JT}: the oxygen-octahedron rotations
shift oxygen atoms from their centrosymmetric positions, so the displacements of oxygen atoms in the JT distortion
do not point directly toward Mn atoms. This reduces the electrostatic repulsion between Mn and oxygen, and therefore favors the JT
distortion. In particular, the calculated Q$_3$ JT distortion in the cubic structure is 0.37 a.u. (see Figure ~\ref{JT}), 
but, it is 0.83 a.u. in the
$Pbnm$ structure (Table ~\ref{alternative}), comparable to the experimentally observed value Q$_3$=0.78 a.u.

We can use the results from Figures~\ref{rotation} and \ref{JT} to see how the coupling of the JT distortion and
rotations is responsible for the stabilization of the $A$-AFM bulk ground state relative to the FM-$M$ phase. In Figure \ref{JT}, we see that 
JT distortion decreases the energy of the $A$-AFM phase. For zero rotational distortion, this energy difference is not big enough to overcome the magnetic energy difference. However, in both phases rotational distortions lead to large energy gains (Figure \ref{rotation}), keeping the energy difference approximately constant. With increased rotations, the energy gain by JT distortion in the $A$-AFM phase is greatly increased, to the point that it is competitive with the magnetic energy difference; a full calculation including relaxation of the lattice parameters then shows that the $A$-AFM phase is favored by 13 meV/fu, as discussed above. We can also understand that it is the Jahn-Teller distortion that opens up the band gap as shown in Fig.~\ref{JT} for $A$-AFM ordering.

The coupling between the JT distortion and the rotational distortions is also evident in Figure~\ref{property} (a) and (b). 
In $Pnma$ systems without a large JT distortion, typically the M rotation decreases and the R rotation increases with increasing tensile strain; in particular, this can be seen here for the FM-$M$ phase below 1\%.
In the $A$-AFM phase, both the M and R rotations increase with increasing tensile strain; the increase in M is the result of coupling to the JT distortion, which is also increasing linearly.

The linear increase in the Jahn-Teller distortion amplitude with tensile strain is directly related to the band gap in the AFM-$I$ phase; a linear increase in the direct band gap from 1.0 eV to 1.3 eV and a linear increase in the indirect band gap from 0.6 eV to 1.0 eV can be seen in Figure \ref{property}(c).
This epitaxial strain control of the band gap, as well as of other features of the electronic structure such as band edges, offers the
opportunity for improving performance for catalytic applications \cite{Berger2011}.
It is already well known that LaMnO$_3$ and other TM oxides such as LaNiO$_3$, with partial electron occupation in the $e_g$ orbital,
exhibit catalytic effects for fuel cells due to their ability to exchange electrons with adsorbed molecules \cite{ORR11}.
Here, we see that tensile epitaxial strain could be used as a way to
shift the $d$-band center relative to the Fermi level and to optimize the subsequent catalytic effects, warranting further investigation
of the electronic structure and catalytic effectiveness of epitaxially-strained LaMnO$_3$ films.

Recent work \cite{film,film2} on the synthesis of LaMnO$_3$ thin films in compressive strain on SrTiO$_3$ substrates 
has shown indications
of a non-bulk ferromagnetic phase which, however, is reported to be insulating, rather than metallic as predicted by our analysis. In our
first-principles investigation, there is no evidence for an insulating FM phase for stoichiometic LaMnO$_3$. This may be due to the limitations of DFT+U in describing the transport properties of correlated-electron materials. However,  ferromagnetic
insulating oxides are known to be quite rare, and another possibility is that the observed phase is in fact the predicted FM-$M$ phase, with the reason for the observed insulating character remaining to be clarified: 
it may be the result of deviations from stoichiometry in the film, or of defects that trap
the free carriers.

The use of square-lattice epitaxial strain to favor a low-orthorhombicity low-energy alternative state over a strongly orthorhombic ground
state could be a promising avenue to pursue in other perovskite oxides.
Other rare-earth $Pnma$ AFM-$I$ manganites are of particular interest, as they are characterized by very large orthorhombicity:
AFM-$I$ $d^4$ $R$MnO$_3$  ($R$=Dy, Ho, Y, Er, Tm, Yb, Lu) have $\frac{b}{a}$-1 $\approx$ 11\% \cite{ortho1,ortho2},
and $R$= Pr, Nd, and Tb have 7\%, 8\%, and 10\% \cite{ortho2}.

In conclusion, we have used first principles calculations for LaMnO$_3$ to investigate the energetics of oxygen-octahedron rotations,
Jahn-Teller distortion, magnetic ordering, and the couplings among them, as a function of epitaxial strain.
This allows us to understand the competition between the ground state $A$-AFM phase and the low-energy alternative FM-$M$ phase, and how
epitaxial strain stabilizes a variety of phases with functional behavior at the phase boundaries.

We would like to thank S.-W. Cheong, D. R. Hamann and D. Vanderbilt for valuable discussions. 
This work was supported by MURI--ARO Grant 
W911NF-07-1-0410, ONR Grant N00014-09-1-0302 and ONR Grant N00014-12-1-1040. 
We acknowledge support from the Center for Scientific Computing at the CNSI and MRL: an NSF MRSEC (DMR-1121053) and NSF CNS-0960316.
K. M. R. thanks the Aspen Center for Physics (NSF Grant 
1066293) where part of this work was carried out.


\begin{thebibliography}{99}

\bibitem{CMR1} P. Schiffer, A. P. Ramirez, W. Bao, S. W. Cheong,  Phys. Rev. Lett. {\bf 75}, 3336 (1995).
\bibitem{CMR2} A. J. Millis, B. I. Shraiman, and R. Mueller, Phys. Rev. Lett. {\bf 77}, 175 (1996). 
\bibitem{CMR3} J. M. D. Coey, M. Viret and S. von Molnar, Adv. Phys. {\bf 48}, 167 (1999).
\bibitem{exp1} Y. Murakami {\it et al}., Phys. Rev. Lett. {\bf 81}, 582 (1998). 
\bibitem{exp2} M. N. Iliev {\it et al}., Phys. Rev. B {\bf 57}, 2872 (1998). 
\bibitem{easy-axis1} F. Moussa {\it et al}., Phys. Rev. B {\bf 54}, 15149 (1996). 
\bibitem{exp4} J. Rodriguez-Carvajal {\it et al}., Phys. Rev. B {\bf 57}, R3189 (1998).
\bibitem{theo1} I. Solovyev, N. Hamada, and K. Terakura, Phys. Rev. B {\bf 53}, 7158 (1996).
\bibitem{easy-axis2} I. Solovyev, N. Hamada, and K. Terakura, Phys. Rev. Lett. {\bf 76}, 4825 (1996). 
\bibitem{failure} H. Sawada, Y. Morikawa, K. Terakura, and N. Hamada, Phys. Rev. B {\bf 56}, 12154  (1997);
                  H. Sawada, Y. Morikawa, N. Hamada, and K. Terakura, J. Magn. Magn. Mater. {\bf 177-181}, 879 (1998);
                  H. Sawada and K. Terakura, Phys. Rev. B {\bf 58}, 6831 (1998); 
                  E. A. Kotomin, R. A. Evarestov, Y. A. Mastrikov, and J. Maier, Phys. Chem. Chem. Phys. {\bf 7}, 2346 (2005).
\bibitem{Leonov} I. Leonov, Dm. Korotin, N. Binggeli, V. I. Anisimov, and D. Vollhardt, Phys. Rev. B {\bf 81}, 075109 (2010).
\bibitem{Terakura10} T. Hashimoto, S. Ishibashi, and K. Terakura, Phys. Rev. B {\bf 82}, 045124 (2010).
\bibitem{JT-Tc} L. Martin-Carron and A. de Andres, The European Physical Journal B, {\bf 22}, 11 (2001).
\bibitem{Billinge} X. Qiu, Th. Proffen, J. F. Mitchell, and S. J. L. Billinge, Phys. Rev. Lett. {\bf 94}, 177203 (2005).
\bibitem{Rondinelli} J. R. Rondinelli and N. A. Spaldin, Adv. Mat. {\bf 23}, 3363 (2011).
\bibitem{Rabe06} C. J. Fennie and K. M. Rabe, Phys. Rev. Lett. {\bf 97}, 267602 (2006).
\bibitem{EuTiO2} J. H. Lee {\it et al}., Nature {\bf 466}, 954 (2010).
\bibitem{ours-SMO} J. H. Lee and K. M. Rabe, Phys. Rev. Lett. {\bf 104}, 207204 (2010). 
\bibitem{ours-SCO} J. H. Lee and K. M. Rabe, Phys. Rev. Lett. {\bf 107}, 067601 (2011). 
\bibitem{uniax1} K. H. Ahn and A. J. Millis, Phys. Rev. B {\bf 64}, 115103 (2001).  
\bibitem{uniax2} B. R. K. Nanda and S. Satpathy, Phys. Rev. B {\bf 81}, 174423 (2010).  
\bibitem{uniax3} B. R. K. Nanda and S. Satpathy, J. Magn. Magn. Mater. {\bf 322}, 3653 (2010).  
\bibitem{uniax4} A. Baena, L. Brey, and M. J. Calderon, Phys. Rev. B {\bf 83}, 064424 (2011). 
\bibitem{coupling} M. A. Carpenter and C. J. Howard,  Acta Cryst. B {\bf 65}, 147 (2009).
\bibitem{mag}J. B. A. A. Elemans {\it et al}., J. Solid State Chem. {\bf 3}, 238 (1971)
\bibitem{GGAU} C. Loschen, J. Carrasco, K. M. Neyman, and F. Illas, Phys. Rev. B {\bf 75}, 035115 (2007).
\bibitem{PBE} J. P. Perdew, K. Burke, and M. Ernzerhof, Phys. Rev. Lett. {\bf 77}, 3865 (1996).
\bibitem{Kresse2} G. Kresse and J. Hafner, Phys. Rev. B {\bf 47}, 558 (1993).
\bibitem{Kresse3} G. Kresse and J. Furthm\"{u}ller, Phys. Rev. B {\bf 54}, 11169 (1996).
\bibitem{Liech} A. I. Liechtenstein, V. I. Anisimov, and J. Zaanen, Phys. Rev. B {\bf 52}, R5467 (1995).
\bibitem{Kresse1} P. E. Bl\"{o}chl, Phys. Rev. B {\bf 50}, 17953 (1994); G. Kresse and D. Joubert, Phys. Rev. B {\bf 59}, 1758 (1999).
\bibitem{strain} N. A. Pertsev, A. G. Zembilgotov, and A. K. Tagantsev, Phys. Rev. Lett. {\bf 80}, 1988 (1998);
                 O. Dieguez et al., Phys. Rev. B 72{\bf 72}, 144101 (2005). 
\bibitem{sro} A. T. Zayak, X. Huang, J. B. Neaton, and K. M. Rabe, Phys. Rev. B {\bf 74}, 094104 (2006).
\bibitem{ab-Pbnm} C.-J. Eklund C. J. Fennie, and K. M. Rabe, Phys. Rev. B {\bf 79}, 220101(R) (2009).
\bibitem{gap1} R. Mahendiran {\it et al}., Appl. Phys. Lett. {\bf 66}, 233 (1995).
\bibitem{Pbnm} S. Coh {\it et al}., Phys. Rev. B {\bf 82}, 064101 (2010).
\bibitem{rotation1} A. J. Hatt and N. A. Spaldin, Phys. Rev. B {\bf 82}, 195402 (2010).  
\bibitem{rotation2} J. M. Rondinelli and S. Coh, Phys. Rev. Lett. {\bf 106}, 235502 (2011).
\bibitem{CMO} S. Bhattacharjee, E. Bousquet, and P. Ghosez,
Phys. Rev. Lett. 102, 117602 (2009).
\bibitem{easy-axis3} S. Mitsudo {\it et al}., J. Magn. Magn. Mater. {\bf 177}, 877 (1998). 
\bibitem{ISODISTORT} ISOTROPY Software Suite, iso.byu.edu; 
       B. J. Campbell, H. T. Stokes, D. E. Tanner, and D. M. Hatch, J. Appl. Cryst. {\bf 39}, 607 (2006).
\bibitem{Berger2011} R. F. Berger, C. J. Fennie and J. B. Neaton, Phys. Rev. Lett. 107, 146804 (2011).
\bibitem{ORR11} J. Suntivich {\it et al}., Nature Chem. {\bf 3}, 647 (2011). 
\bibitem{gap2}T. Arima, Y. Tokura, and J. B. Torrance, Phys. Rev. B {\bf 48}, 17006 (1993).
\bibitem{film} C. Adamo {\it et al}., Appl. Phys. Lett. {\bf 92}, 112508 (2008).
\bibitem{film2} Z. Marton, S. S. A. Seo, T. Egami and H. N. Lee, J. Crystal Growth 312, 2923 (2010). 
\bibitem{ortho1} M. Tachibana {\it et al}., Phys. Rev. B {\bf 75}, 144425 (2007).  
\bibitem{ortho2} J. A. Alonso, M. J. Martinez-Lope, M. T. Casais, and M. T. Fernandez-Diaz, Inorg. Chem. {\bf 39}, 917 (2000).  
\bibitem{ortho3} P. Lacorre {\it et al}., J. Solid State. Chem. {\bf 91}, 225 (1991); 
M. T. Escote, A. M. da Silva, J. R. Matos, and R. F. Jardim, 
J.Solid State Chem. {\it 151},  298 (2000).
\end{thebibliography}
\end{document}